\documentclass[%
 reprint,
superscriptaddress,
 amsmath,amssymb,
 aps,
]{revtex4-2}

\usepackage{graphicx}
\usepackage{dcolumn}
\usepackage{bm}

\usepackage{color}
\begin{document}

\preprint{APS/123-QED}

\title{Deformation of hydrogel during freezing}

\author{Lila Séguy}
 \affiliation{%
 Institut Jean Le Rond $\rm \partial$'Alembert, Sorbonne Université, CNRS, UMR 7190, 75005 Paris, France\\
}%

 \author{Axel Huerre}
\affiliation{
Laboratoire Matière et Systèmes Complexes (MSC), Université Paris Cité, CNRS, UMR 7057, 75013 Paris, France\\
}%

\author{Suzie Protière}%
\affiliation{%
 Institut Jean Le Rond $\rm \partial$'Alembert, Sorbonne Université, CNRS, UMR 7190, 75005 Paris, France\\
}%


\date{\today}

\begin{abstract}
We conducted experiments to directionally freeze agar hydrogel drops with different polymer concentrations, on a copper substrate maintained at low temperature. Unlike the water droplet studied in the literature, where the liquid part can reorganize during freezing, our droplets show a distinct deformation. This is due to the presence of the viscoelastic solid polymer matrix, which prevents deformation of the unfrozen part, leading mainly to elongation in the direction of the temperature gradient. We then derive a model describing this deformation. It assumes a flat freezing front, a change in volume at the front resulting from the Stefan condition, and a remaining volume to be frozen identical in shape to the initial one. The latter is in good agreement with experimental observations as long as water fluxes in the porous hydrogel remain negligible. It is also conclusive for different object shapes, as the hydrogel can be freely molded without a container.
\end{abstract}

\maketitle




Upon freezing, water undergoes a volume increase of around 9\% due to the difference in density between water and ice ($\rho_{w} = 997$ kg.m$^{-3}$, $\rho_{i} = 917$ kg.m$^{-3}$). At the capillary scale, this effect leads to radical changes in the shape of the frozen object, for instance with the formation of a surprising tip on the top of a frozen water droplet \cite{Stairs1971a,Anderson1996a,Snoeijer2012b,Schetnikov2015a,Nauenberg2016,Marin2014a}.
When aqueous materials organized at different scales are subjected to freezing, their properties can also be altered. This is the case for example with suspensions where the frozen ice can engulf the particles or reject them \cite{Peppin2006}, leading to the appearance of ice-lenses and water network reorganization when freezing a soil \cite{zotero-1005,Rempel2007}. 
Hydrogels typically present a small-scale organization. They are composed of a three-dimensional matrix of chemically or physically cross-linked polymers, containing a significant proportion of water, sometimes close to 100\% of its mass. When this material freezes, the ice can reorganize the bonds between the polymers, modifying the porosity of the material \cite{Grenier2019}. 
Using this property, ice-templating methods developed \cite{Deville2006}, where a fine control on the ice properties can be used to achieve specific porosity in the final material.
Ice-templated porous hydrogels have been successfully synthesized for applications such as tissue reconstruction \cite{Nugent2005, DeFrance2018} and food preservation \cite{Khalesi2020,Xu2016}.
In this experimental and theoretical study, we show that the freezing of hydrogel samples induces a non-homogeneous dilatation of the material. The deformation is localized in the direction of the phase change and depends on the temperature of freezing and on the proportion of polymers. A model of the phase change including the unique ability of hydrogels to retain a signature of their original shape allows for the good prediction of the frozen shape of the material for any initial shape.



\begin{figure}[!]
\includegraphics{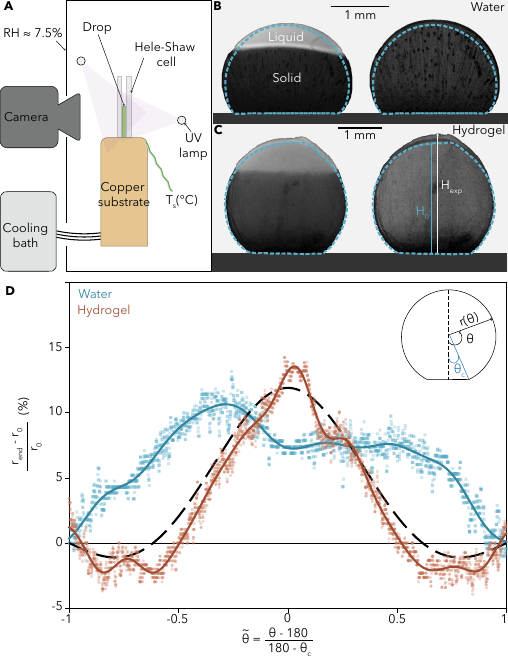}
\caption{\label{fig1} (a) Schematics of the experimental setup.
(b) Snapshots of a freezing drop of mixture of water and fluorescein ($0.5$ g/L) on a cold substrate at $T_{s} = -11^{\circ}$C. 
(c) Snapshots of a 4\% freezing hydrogel drop on a cold substrate at $T_{s} = -37^{\circ}$C. 
The initial shape is represented with a dashed blue line.
(d) Relative change in the local radius between the totally frozen drop ($\rm r_{\rm end}$) and the initial drop ($r_0$) as a function of a normalized angle $\rm \tilde\theta = \frac{\theta - 180}{180 - \theta_{c}}$. The geometrical parameters are described in the inset. Raw data are plotted as points for water (blue) and hydrogel (red) and the curves are obtained from a low-pass filter smoothing. The black dashed curve corresponds to the analytical model for the hydrogel drop (eq. \ref{eq3}).}
\end{figure}


Hydrogel samples with various shapes were synthesized by dissolving agar (CAS 9002-18-0) in a a  mixture of water and fluorescein (0.5 g/L). 
The proportion of polymer was fixed to $ C = \frac{m_{agar}}{m_{water}} = 0.5,1,2,3,4,5 \%$ giving an elastic modulus between $\rm E_{\infty} \sim 0.6$ kPa for $C = \rm 0.5 \%$ and $\rm E_{\infty} \sim 50$ kPa for $C$ = 4$\%$
The solution is then covered and maintained at $70^{\circ}$C to obtain a homogeneous mixture. 
A mold of the desired shape, made out of PolyVynilSiloxane (PVS Zhermak Elite Double 22), is placed between two silanized glass plates spaced 200 $\mu$m apart.
The mixture is poured in the mold and gelifies within seconds. After removing the mold, we obtain a hydrogel sample of specific shape and composition sandwiched between two glass slides forming a Hele-Shaw cell.
The sample is finally put in contact with a cold copper substrate (temperature $T_{s} = -10^{\circ}$C to $-50 ^{\circ}$C) to induce freezing, as schematized on Fig. \ref{fig1}(a).
The whole experiment is placed in a low-humidity box to prevent the formation of frost.
The sample is lighted with UV lamps to obtain a significant contrast between the ice and the fluorescent liquid \cite{Monier2020,Seguy2023a,Bumma2023}.
The progression of the freezing front, from bottom to top, is recorded from the side at 25 fps using a Nikon D800 camera equipped with a Navitar x12 lens. 
Contour detection is then performed on the pictures to extract the shape of the sample at each time step.
Typical pictures obtained from the experiment are presented for a drop of MiilliQ water (Fig. \ref{fig1}(b), $T_s = -11^{\circ}$C) and for a drop of hydrogel (Fig. \ref{fig1}(c), $C = 4\%, T_s = -37^{\circ}$C). 
For each example, the picture on the left is taken at t = 35 s and clearly shows the position of the solidification front. The picture on the right shows the final frozen shape with the initial shape (dashed blue line) superimposed to highlight the deformation of the sample and the qualitative difference between water and hydrogel.


Indeed, the presence of the polymer matrix influences the final frozen shape as the direction of deformation of the drop is mostly localized with a contraction of the base of the hydrogel drop, a maximal width that remains mostly unchanged and a larger increase of the height of the material(see also corresponding videos of the process in Supp. Mat.).

We must note that, contrary to the 3D case \cite{Stairs1971a}, we do not observe a tip at the top of the drop on Fig. \ref{fig1}(b). This could be due to the flatness of the freezing front as the classic mechanism for the formation of the tip is based on a circular front. 
In 3D, the negligible heat flux in the air is invoked to justify that the solidification front must locally be orthogonal to the vapor-water interface. However, here, due to the presence of the cell walls and the resulting 2D geometry this assumption breaks down and allows for a flat front. Increasing the distance between the cell walls of the Hele-Shaw cell mitigates this effect as we observe a tip for plates that are 1 mm apart (Supp. Mat.), similarly to 3D studies \cite{Marin2014a}. 


We can quantify these differences in shape by plotting the relative radius change along the surface for both drops in Fig. \ref{fig1}(b) and (c), as a function of the normalized angle $\rm \tilde\theta = \frac{\theta - 180}{180 - \theta_{c}}$ (inset Fig. \ref{fig1}(d)) ranging between -1 and 1. $\theta_{c}$ corresponds to the left and right bottom limits of the drop. 

The water drop (blue curve, Fig. \ref{fig1}(d)) seems to deform uniformly with an overall deformation along the surface lying between 5\% and 10\%. This confirms an anisotropic distribution of the volume change in the case of the water drop. 
However, for the hydrogel drop (red curve, Fig. \ref{fig1}(d)), we find negative values of the deformation for $\rm \tilde\theta < -0.5$ and $\rm \tilde\theta > 0.5$. This means that the hydrogel contracts close to its area of contact with the copper substrate. Moreover, a sharp peak is observed at the top of the hydrogel drop for $\rm \tilde\theta = 0$ with around 15\% change in local radius, thus highlighting that a strong expansion is localized at the apex after freezing. 
In order to assess globally the effect of the agar matrix on the solidification process, we compute the overall volume increase for all the experiments.

\begin{figure}[!]
\includegraphics{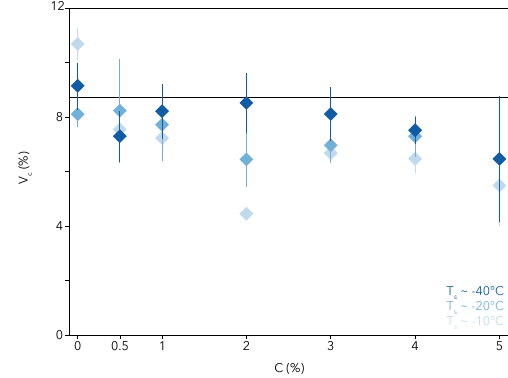}
\caption{\label{fig2} Mean volume change $\rm v_{c}$ for different polymer concentrations at different substrate temperatures. The black line represents the theoretical value for pure water ($v_{c} = 9.2\%$).}
\end{figure}

\begin{figure*}[!]
\includegraphics{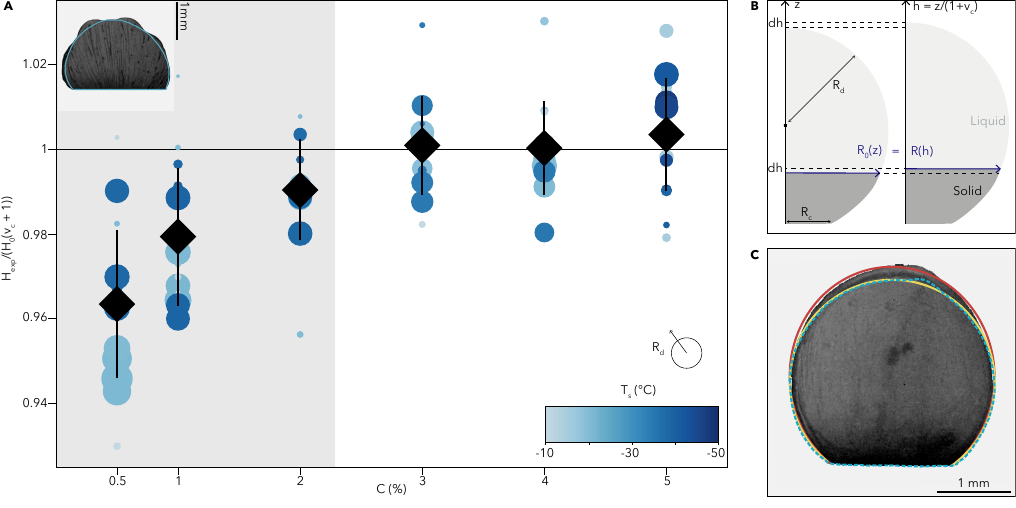}
\caption{\label{fig3} (a) $H_{\rm exp}/(H_{0}(v_{c}+1))$ for different polymer concentrations, different temperatures and different sizes. Blue points are colored accordingly to the temperature and black ones are the mean value for each concentration. The size of the blue points is linked to the drop radius. The gray area delimits experiments where a significant water release is observed and the model is expected to fail. The horizontal black line corresponds to the situation where the volume change is only vertical. In this case we usually observe frozen shapes such as the one presented in inset (upper left picture with a 0.5\% drop on a substrate at $T_{s} = -11^{\circ}C$). (b) Schematics of the model: the liquid part is in light gray and the frozen part in dark gray. The contact area is of radius $2R_{c}$ and the insitial shape is described by the function $R_{0}(z)$. The freezing front is at height $h(t)$ and has a radius $R(t)$. (c) Frozen agar drop on a substrate at $T_{s} = -37^{\circ}$C and $C=4\%$. The blue dashed line corresponds to the initial shape and the yellow one its circle approximation. The red line represents the final theoretical contour for hydrogel (equation \ref{eq3}).}
\end{figure*}



This volume increase, $v_{c}$, is presented on Figure \ref{fig2} as a function of the polymer concentration, for three different substrate temperatures. 
It is derived from the integral of curves such as the one presented in Fig. \ref{fig1} (d).
The black line represents the expected value of 9.2\% for the volume increase when water is changed to ice. The first points at $C = 0\%$ were obtained with pure water drops and agree with the expected value.

When the concentration of polymers in the hydrogel is increased, the volume change due to freezing of the sample decreases. Indeed, for a 5\% hydrogel, around 3\% in volume change is lost. 
We also find that for concentrations above 1\%, the volume change is increasing when the substrate temperature decreases: the dark blue points (-40°C) consistently show a greater expansion than samples at higher temperatures.
This decrease in volume change for hydrogels could be linked to the volume of concentrated polymer solution that cannot freeze.
First, the presence of the polymer destabilizes the freezing front in crystals aligned in the direction of the temperature gradient \cite{Mullins1964}. Between these crystals an interstitial liquid, composed of water highly concentrated in polymer, is present \cite{Langer1980, Worster1986, Huppert1990}.
The increasing concentration of solute induces a depression of the freezing point \cite{Banin1974}, ultimately preventing the freezing and explaining why the volume change in the case of a hydrogel is lower than for water.

Attempts to predict the dependence of the volume change with both the temperature and the concentration theoretically are presented in Supp. Mat. Either based on the Clausius-Clapeyron relation or on empirical laws \cite{Fikiin1998, Succar1990, Rahman2009}, the predictions give a reasonable trend but the agreement failed to be quantitative. We will thus use the experimental points of Figure \ref{fig2} as a reference for the rest of the study.

We now wish to investigate how this global volume variation is related to a local vertical expansion for a frozen hydrogel drop.
We therefore measure the initial and final height ($H_{\rm 0}$ and $H_{\rm exp}$ respectively) for hydrogel drops of various concentrations and temperatures. We can then plot the ratio between the drop's experimental final height and $H_{0}(1+v_{c}(T_{s},C))$, the final height of a droplet if the volume change were purely unidirectional (Figure \ref{fig3}(a)), as a function of the hydrogel concentration. Here, $v_{c}(T_{s}, C)$ is the mean volume change for a given temperature and concentration measured in Fig. \ref{fig2}.

The blue data points represent the experimental data at varying temperature and the black ones the mean values for each concentration. The size of the dot is proportional to the radius of the drop $R_d$.
Interestingly, for $C\,>\,2\%$ the height change is close to the total volume change ($\rm \frac{H_{\rm exp}}{H_{0}} = v_{c} + 1$), all of the deformation is quantitatively handled in one and only direction: perpendicular to the solidification front. The small error mainly comes from the irregularities at the surface of the drop and from the experimental reproducibility. 

For concentrations below $2\%$, the vertical ascent of the freezing front is below the predicted value, especially for 0.5\% and 1\%. 
Indeed, in this case, water is released at the surface of the hydrogel during the freezing process as presented in inset of Figure \ref{fig3} (and corresponding movie in Supp. Mat). This water release leads to the formation of bumps at random positions along the surface of the frozen hydrogel, preventing the volume change from being fully vertical. 
The poroviscoelastic nature of the hydrogel \cite{Caccavo2017, Caccavo2018, Hu2012} could explain that a stress stemming from the volume increase could trigger a flow of the water in the polymer matrix and its subsequent release along the drop's surface. The dynamics of the water migration would then depend on the size and porosity of the sample and would be much faster for low-concentrated hydrogels as the pore size would then be larger \cite{Davies2010}.

The classic theoretical description for the freezing of a drop of water relies on the ability of the liquid part to rearrange its shape through capillary forces \cite{Anderson1996a} (see Supp. Mat. for a model describing the 2D case and comparison with experimental results). 
However, in the case of a hydrogel, although it almost entirely consists of water, it has a unique behavior with the water being maintained by the polymer matrix and thus preventing rearrangements along the freezing front. We can thus assume that the elastic nature of this soft solid allows the shape of the unfrozen part to remain constant throughout the solidification process. We propose a theoretical model where the deformation is then entirely localized at the front where the local height is subjected to an expansion only in the vertical direction. This deformation spreads towards the top during the freezing process similarly to a propagation front observed in a phase-transforming material such as kirigami shells under strain \cite{Rafsanjani2019}.
 

First, let us consider a ($r$,$z$)-surface with a central axis of symmetry parameterized with its distance from the axis $R$ as a function of the position on this axis $z$ so that for each $z$, $R$ is defined without ambiguity. 
The initial shape, of volume $v_0$, is described by the function $R_0(z)$, $z$ varying from 0 to $H_0$. 

When a layer of hydrogel of thickness $dz$, initially at the position $z$, is changed into ice, the front progresses of $dh=(1+v_{c})dz$, as illustrated on the schematics Figure \ref{fig3}(b).
This leads to an explicit expression for the final shape of the frozen object $R(z)$:

\begin{equation}
    R(z)=R_0(\frac{z}{1+v_c})
\label{eq:expension}
\end{equation}
with $z$ now varying from $0$ to $H_0(1+v_c)$.
From this model, we can reconstruct the theoretical frozen shape of any initial hydrogel shape.


For the case of a droplet, modeled as a portion of disk of radius $\rm R_{d}$ in contact with the substrate over a distance $2R_{c}$ (Fig. \ref{fig3}(b)), we can derive an analytical expression for the frozen drop.
It is given by:

\begin{equation}
    \left\{
        \begin{array}{ll}
            H(R) = (1+v_c)(\sqrt{R_{d}^2 - R_{c}^2}-\sqrt{R_{d}^2 - R^2}) \\
            H(R) = (1+v_c)(\sqrt{R_{d}^2 - R_{c}^2} +\sqrt{R_{d}^2 - R^2}),
        \end{array}
    \right.
    \label{eq3}
\end{equation}
with $H$ the height of the object at a given horizontal radius. The first line of equation (\ref{eq3}) corresponds to a portion of ellipse below its equator and the second line above. Thus, a liquid truncated disk of radius $R_{d}$ will give a frozen truncated ellipse of horizontal axis $R_{d}$ and vertical axis $(1+v_c)R_{d}$. Indeed, our condition on the conservation of the initial shape does not allow any change of the liquid portion in another direction than the freezing front, leading to an unchanged horizontal axis.

Figure \ref{fig3}(c) presents how our model can predict the final shape of an experimental agar drop corresponding to Figure \ref{fig1}(b). The initial drop shape (blue dashed line) is fitted with a circle (yellow line) and the final ellipse is in red. We find a good agreement between theory and experiments, the only measurable error comes from the lost of circularity of the frozen drop that is reminiscent from the same defect in the initial state.
Another way of confronting the model to our experimental results is to plot the variation of radius before and after freezing along the angle $\theta$, as discussed previously in Figure \ref{fig1}(d). The theoretical radius deformation of this particular experiment is plotted in black dashed line. The elliptical shape reproduces both of our observations: a negative change in radius close to the base of the drop and a maximal deformation at its top. 


The analytical approach can be further extended to other shapes, such as a rectangle or a triangle.
For the rectangular case, described by $R_0(z)$ = $R_c$ for $0 \leq  z \leq H_0$, the frozen shape is given by $R(z)=R_c$ but for for $0 \leq  z \leq H_0(1+v_c)$. The initial rectangle is dilated in the vertical direction.
Similarly, for an initial triangular shape with a top angle $2\alpha$ and a base $2R_c$, $R(z)=R_c - z {\rm tan}(\alpha) $, the frozen shape is also a triangle with a base $2R_c$ but with a top angle $\beta = {\rm arctan}({\rm tan}(\alpha)/(1+v_c))$.
For a given initial shape with small defects, equation (\ref{eq:expension}) can be used numerically to predict the frozen pattern.

\begin{figure}[!]
\includegraphics{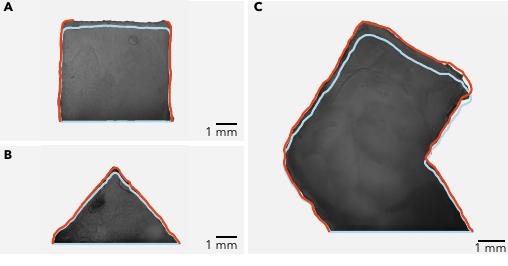}
\caption{\label{fig4} Frozen hydrogel shapes (2\%) (a) Square on a substrate at $T_{s} = -57^{\circ}$C (b) Triangle on a substrate at $T_{s} = -86^{\circ}$C (c) Asymmetric shape on a substrate at $T_{s} = -97^{\circ}$C. Initial shapes are represented by the blue lines and the final theoretical ones by the red lines.}
\end{figure}

Figure \ref{fig4} shows different frozen shapes with the initial contour in blue and the theoretical deformation in red. We tested different evolution of horizontal section: constant (Fig. \ref{fig4}(a)), constantly decreasing (Fig. \ref{fig4}(b)) and non monotonic (Fig. \ref{fig4}(c)). Experiments were done on 2\% hydrogels with a substrate temperature below $-50^{\circ}$C, preventing any water release. 
As predicted above, we observe that the rectangle on Fig. \ref{fig4}(a) has not gained in width but just expanded vertically. 
This is not the case for the triangle on Fig. \ref{fig4}(b) where the final shape, for theory and experiment, is larger than the initial one. As expected, the surface is vertically lifted up but is also keeping is base's length giving a new triangle with a smaller angle at the top. These experimental observations remain valid for experiments with shapes where a brutal change of monotony is imposed on the horizontal section. Figure \ref{fig4}(c) shows a frozen asymmetric hydrogel that has opposite monotony on both sides.
On the right, the frozen part has slightly dilated after freezing on the lower half part and then has contracted in the upper half after the change in evolution of the section. The opposite phenomenon is observed on the left side which has two changes in section monotony.
Interestingly, most of the deformation is located on the top part of the object, as the vertical deformation is proportional to the height of the object. Overall, our model predicts well the shape of frozen hydrogels.


In this study, we present a theoretical model that describes quantitatively the deformation of hydrogels during freezing. Unlike a drop of water that can rearrange its shape during the process, the viscoelastic polymer matrix will maintain the drop's initial shape, leading to the propagation of a vertical deformation of the object only in the direction of the phase transition. We also investigated the range of validity of our model and showed that it works at low substrate temperature on concentrated hydrogels, as it prevents significant water movements. Finally, we successfully tested our theory on different hydrogel shapes. This problem could be adapted to 3D shapes by removing the assumption of a flat freezing front \cite{Stiti2020a} which would lead to more complex calculation. Furthermore, the water release phenomenon should be investigated, as it could mean that the material undergoes physical properties changes (water content, porosity) during its freezing.

\subsection*{Acknowledgements}
SP would like to thank Andreas Carlson who contributed to preliminary stages of this work during a visit at d'Alembert. We thank L. Duchemin for fruitful discussions about the modeling.

\bibliography{article_defo_hydro.bib}

\begin{thebibliography}{32}%
\makeatletter
\providecommand \@ifxundefined [1]{%
 \@ifx{#1\undefined}
}%
\providecommand \@ifnum [1]{%
 \ifnum #1\expandafter \@firstoftwo
 \else \expandafter \@secondoftwo
 \fi
}%
\providecommand \@ifx [1]{%
 \ifx #1\expandafter \@firstoftwo
 \else \expandafter \@secondoftwo
 \fi
}%
\providecommand \natexlab [1]{#1}%
\providecommand \enquote  [1]{``#1''}%
\providecommand \bibnamefont  [1]{#1}%
\providecommand \bibfnamefont [1]{#1}%
\providecommand \citenamefont [1]{#1}%
\providecommand \href@noop [0]{\@secondoftwo}%
\providecommand \href [0]{\begingroup \@sanitize@url \@href}%
\providecommand \@href[1]{\@@startlink{#1}\@@href}%
\providecommand \@@href[1]{\endgroup#1\@@endlink}%
\providecommand \@sanitize@url [0]{\catcode `\\12\catcode `\$12\catcode
  `\&12\catcode `\#12\catcode `\^12\catcode `\_12\catcode `\%12\relax}%
\providecommand \@@startlink[1]{}%
\providecommand \@@endlink[0]{}%
\providecommand \url  [0]{\begingroup\@sanitize@url \@url }%
\providecommand \@url [1]{\endgroup\@href {#1}{\urlprefix }}%
\providecommand \urlprefix  [0]{URL }%
\providecommand \Eprint [0]{\href }%
\providecommand \doibase [0]{https://doi.org/}%
\providecommand \selectlanguage [0]{\@gobble}%
\providecommand \bibinfo  [0]{\@secondoftwo}%
\providecommand \bibfield  [0]{\@secondoftwo}%
\providecommand \translation [1]{[#1]}%
\providecommand \BibitemOpen [0]{}%
\providecommand \bibitemStop [0]{}%
\providecommand \bibitemNoStop [0]{.\EOS\space}%
\providecommand \EOS [0]{\spacefactor3000\relax}%
\providecommand \BibitemShut  [1]{\csname bibitem#1\endcsname}%
\let\auto@bib@innerbib\@empty
\bibitem [{\citenamefont {Stairs}()}]{Stairs1971a}%
  \BibitemOpen
  \bibfield  {author} {\bibinfo {author} {\bibfnamefont {R.~A.}\ \bibnamefont
  {Stairs}},\ }\bibfield  {title} {\bibinfo {title} {Changes of drop-shapes on
  freezing},\ }\href {https://doi.org/10.1021/ac60305a036} {\ \textbf {\bibinfo
  {volume} {43}},\ \bibinfo {pages} {1535}}\BibitemShut {NoStop}%
\bibitem [{\citenamefont {Anderson}\ \emph {et~al.}()\citenamefont {Anderson},
  \citenamefont {Worster},\ and\ \citenamefont {Davis}}]{Anderson1996a}%
  \BibitemOpen
  \bibfield  {author} {\bibinfo {author} {\bibfnamefont {D.~M.}\ \bibnamefont
  {Anderson}}, \bibinfo {author} {\bibfnamefont {M.~G.}\ \bibnamefont
  {Worster}},\ and\ \bibinfo {author} {\bibfnamefont {S.~H.}\ \bibnamefont
  {Davis}},\ }\bibfield  {title} {\bibinfo {title} {The case for a dynamic
  contact angle in containerless solidification},\ }\href
  {https://doi.org/10.1016/0022-0248(95)00970-1} {\ \textbf {\bibinfo {volume}
  {163}},\ \bibinfo {pages} {329}}\BibitemShut {NoStop}%
\bibitem [{\citenamefont {Snoeijer}\ and\ \citenamefont
  {Brunet}()}]{Snoeijer2012b}%
  \BibitemOpen
  \bibfield  {author} {\bibinfo {author} {\bibfnamefont {J.~H.}\ \bibnamefont
  {Snoeijer}}\ and\ \bibinfo {author} {\bibfnamefont {P.}~\bibnamefont
  {Brunet}},\ }\bibfield  {title} {\bibinfo {title} {Pointy ice-drops: {{How}}
  water freezes into a singular shape},\ }\href
  {https://doi.org/10.1119/1.4726201} {\ \textbf {\bibinfo {volume} {80}},\
  \bibinfo {pages} {764}}\BibitemShut {NoStop}%
\bibitem [{\citenamefont {Schetnikov}\ \emph {et~al.}()\citenamefont
  {Schetnikov}, \citenamefont {Matiunin},\ and\ \citenamefont
  {Chernov}}]{Schetnikov2015a}%
  \BibitemOpen
  \bibfield  {author} {\bibinfo {author} {\bibfnamefont {A.}~\bibnamefont
  {Schetnikov}}, \bibinfo {author} {\bibfnamefont {V.}~\bibnamefont
  {Matiunin}},\ and\ \bibinfo {author} {\bibfnamefont {V.}~\bibnamefont
  {Chernov}},\ }\bibfield  {title} {\bibinfo {title} {Conical shape of frozen
  water droplets},\ }\href {https://doi.org/10.1119/1.4897499} {\ \textbf
  {\bibinfo {volume} {83}},\ \bibinfo {pages} {36}}\BibitemShut {NoStop}%
\bibitem [{\citenamefont {Nauenberg}()}]{Nauenberg2016}%
  \BibitemOpen
  \bibfield  {author} {\bibinfo {author} {\bibfnamefont {M.}~\bibnamefont
  {Nauenberg}},\ }\bibfield  {title} {\bibinfo {title} {Theory and experiments
  on the ice–water front propagation in droplets freezing on a subzero
  surface},\ }\href {https://doi.org/10.1088/0143-0807/37/4/045102} {\ \textbf
  {\bibinfo {volume} {37}},\ \bibinfo {pages} {045102}}\BibitemShut {NoStop}%
\bibitem [{\citenamefont {Marín}\ \emph {et~al.}()\citenamefont {Marín},
  \citenamefont {Enríquez}, \citenamefont {Brunet}, \citenamefont {Colinet},\
  and\ \citenamefont {Snoeijer}}]{Marin2014a}%
  \BibitemOpen
  \bibfield  {author} {\bibinfo {author} {\bibfnamefont {A.~G.}\ \bibnamefont
  {Marín}}, \bibinfo {author} {\bibfnamefont {O.~R.}\ \bibnamefont
  {Enríquez}}, \bibinfo {author} {\bibfnamefont {P.}~\bibnamefont {Brunet}},
  \bibinfo {author} {\bibfnamefont {P.}~\bibnamefont {Colinet}},\ and\ \bibinfo
  {author} {\bibfnamefont {J.~H.}\ \bibnamefont {Snoeijer}},\ }\bibfield
  {title} {\bibinfo {title} {Universality of {{Tip Singularity Formation}} in
  {{Freezing Water Drops}}},\ }\href
  {https://doi.org/10.1103/PhysRevLett.113.054301} {\ \textbf {\bibinfo
  {volume} {113}},\ \bibinfo {pages} {054301}}\BibitemShut {NoStop}%
\bibitem [{\citenamefont {Peppin}\ \emph {et~al.}()\citenamefont {Peppin},
  \citenamefont {Elliott},\ and\ \citenamefont {Worster}}]{Peppin2006}%
  \BibitemOpen
  \bibfield  {author} {\bibinfo {author} {\bibfnamefont {S.}~\bibnamefont
  {Peppin}, \bibfnamefont {Stephen}}, \bibinfo {author} {\bibfnamefont
  {J.~a.~W.}\ \bibnamefont {Elliott}},\ and\ \bibinfo {author} {\bibfnamefont
  {M.~G.}\ \bibnamefont {Worster}},\ }\bibfield  {title} {\bibinfo {title}
  {Solidification of colloidal suspensions},\ }\href
  {https://doi.org/10.1098/rspa.2006.1790} {\ \textbf {\bibinfo {volume}
  {554}},\ \bibinfo {pages} {147}}\BibitemShut {NoStop}%
\bibitem [{\citenamefont {Miller}()}]{zotero-1005}%
  \BibitemOpen
  \bibfield  {author} {\bibinfo {author} {\bibfnamefont {R.}~\bibnamefont
  {Miller}},\ }\bibfield  {title} {\bibinfo {title} {{{Freezing and heaving of
  saturated and unsaturated soils}}},\ }\href@noop {} {\ }\BibitemShut
  {NoStop}%
\bibitem [{\citenamefont {Rempel}(2007)}]{Rempel2007}%
  \BibitemOpen
  \bibfield  {author} {\bibinfo {author} {\bibfnamefont {A.~W.}\ \bibnamefont
  {Rempel}},\ }\bibfield  {title} {\bibinfo {title} {Formation of ice lenses
  and frost heave},\ }\bibfield  {journal} {\bibinfo  {journal} {Journal of
  Geophysical Research: Earth Surface}\ }\textbf {\bibinfo {volume} {112}},\
  \href {https://doi.org/10.1029/2006JF000525} {10.1029/2006JF000525} (\bibinfo
  {year} {2007})\BibitemShut {NoStop}%
\bibitem [{\citenamefont {Grenier}\ \emph {et~al.}()\citenamefont {Grenier},
  \citenamefont {Duval}, \citenamefont {Barou}, \citenamefont {Lv},
  \citenamefont {David},\ and\ \citenamefont {Letourneur}}]{Grenier2019}%
  \BibitemOpen
  \bibfield  {author} {\bibinfo {author} {\bibfnamefont {J.}~\bibnamefont
  {Grenier}}, \bibinfo {author} {\bibfnamefont {H.}~\bibnamefont {Duval}},
  \bibinfo {author} {\bibfnamefont {F.}~\bibnamefont {Barou}}, \bibinfo
  {author} {\bibfnamefont {P.}~\bibnamefont {Lv}}, \bibinfo {author}
  {\bibfnamefont {B.}~\bibnamefont {David}},\ and\ \bibinfo {author}
  {\bibfnamefont {D.}~\bibnamefont {Letourneur}},\ }\bibfield  {title}
  {\bibinfo {title} {Mechanisms of pore formation in hydrogel scaffolds
  textured by freeze-drying},\ }\href
  {https://doi.org/10.1016/j.actbio.2019.05.070} {\ \textbf {\bibinfo {volume}
  {94}},\ \bibinfo {pages} {195}}\BibitemShut {NoStop}%
\bibitem [{\citenamefont {Deville}\ \emph {et~al.}(2006)\citenamefont
  {Deville}, \citenamefont {Saiz}, \citenamefont {Nalla},\ and\ \citenamefont
  {Tomsia}}]{Deville2006}%
  \BibitemOpen
  \bibfield  {author} {\bibinfo {author} {\bibfnamefont {S.}~\bibnamefont
  {Deville}}, \bibinfo {author} {\bibfnamefont {E.}~\bibnamefont {Saiz}},
  \bibinfo {author} {\bibfnamefont {R.~K.}\ \bibnamefont {Nalla}},\ and\
  \bibinfo {author} {\bibfnamefont {A.~P.}\ \bibnamefont {Tomsia}},\ }\bibfield
   {title} {\bibinfo {title} {Freezing as a {{Path}} to {{Build Complex
  Composites}}},\ }\href {https://doi.org/10.1126/science.1120937} {\bibfield
  {journal} {\bibinfo  {journal} {Science}\ }\textbf {\bibinfo {volume}
  {311}},\ \bibinfo {pages} {515} (\bibinfo {year} {2006})}\BibitemShut
  {NoStop}%
\bibitem [{\citenamefont {Nugent}\ \emph {et~al.}()\citenamefont {Nugent},
  \citenamefont {Hanley}, \citenamefont {Tomkins},\ and\ \citenamefont
  {Higginbotham}}]{Nugent2005}%
  \BibitemOpen
  \bibfield  {author} {\bibinfo {author} {\bibfnamefont {M.~J.~D.}\
  \bibnamefont {Nugent}}, \bibinfo {author} {\bibfnamefont {A.}~\bibnamefont
  {Hanley}}, \bibinfo {author} {\bibfnamefont {P.~T.}\ \bibnamefont
  {Tomkins}},\ and\ \bibinfo {author} {\bibfnamefont {C.~L.}\ \bibnamefont
  {Higginbotham}},\ }\bibfield  {title} {\bibinfo {title} {Investigation of a
  novel freeze-thaw process for the production of drug delivery hydrogels},\
  }\href {https://doi.org/10.1007/s10856-005-4722-7} {\ \textbf {\bibinfo
  {volume} {16}},\ \bibinfo {pages} {1149}}\BibitemShut {NoStop}%
\bibitem [{\citenamefont {De~France}\ \emph {et~al.}()\citenamefont
  {De~France}, \citenamefont {Xu},\ and\ \citenamefont {Hoare}}]{DeFrance2018}%
  \BibitemOpen
  \bibfield  {author} {\bibinfo {author} {\bibfnamefont {K.~J.}\ \bibnamefont
  {De~France}}, \bibinfo {author} {\bibfnamefont {F.}~\bibnamefont {Xu}},\ and\
  \bibinfo {author} {\bibfnamefont {T.}~\bibnamefont {Hoare}},\ }\bibfield
  {title} {\bibinfo {title} {Structured {{Macroporous Hydrogels}}:
  {{Progress}}, {{Challenges}}, and {{Opportunities}}},\ }\href
  {https://doi.org/10.1002/adhm.201700927} {\ \textbf {\bibinfo {volume} {7}},\
  \bibinfo {pages} {1700927}}\BibitemShut {NoStop}%
\bibitem [{\citenamefont {Khalesi}\ \emph {et~al.}()\citenamefont {Khalesi},
  \citenamefont {Lu}, \citenamefont {Nishinari},\ and\ \citenamefont
  {Fang}}]{Khalesi2020}%
  \BibitemOpen
  \bibfield  {author} {\bibinfo {author} {\bibfnamefont {H.}~\bibnamefont
  {Khalesi}}, \bibinfo {author} {\bibfnamefont {W.}~\bibnamefont {Lu}},
  \bibinfo {author} {\bibfnamefont {K.}~\bibnamefont {Nishinari}},\ and\
  \bibinfo {author} {\bibfnamefont {Y.}~\bibnamefont {Fang}},\ }\bibfield
  {title} {\bibinfo {title} {New insights into food hydrogels with reinforced
  mechanical properties: {{A}} review on innovative strategies},\ }\href
  {https://doi.org/10.1016/j.cis.2020.102278} {\ \textbf {\bibinfo {volume}
  {285}},\ \bibinfo {pages} {102278}}\BibitemShut {NoStop}%
\bibitem [{\citenamefont {Xu}\ \emph {et~al.}()\citenamefont {Xu},
  \citenamefont {Tao},\ and\ \citenamefont {Shivkumar}}]{Xu2016}%
  \BibitemOpen
  \bibfield  {author} {\bibinfo {author} {\bibfnamefont {Y.}~\bibnamefont
  {Xu}}, \bibinfo {author} {\bibfnamefont {Y.}~\bibnamefont {Tao}},\ and\
  \bibinfo {author} {\bibfnamefont {S.}~\bibnamefont {Shivkumar}},\ }\bibfield
  {title} {\bibinfo {title} {Effect of freeze-thaw treatment on the structure
  and texture of soft and firm tofu},\ }\href
  {https://doi.org/10.1016/j.jfoodeng.2016.06.022} {\ \textbf {\bibinfo
  {volume} {190}},\ \bibinfo {pages} {116}}\BibitemShut {NoStop}%
\bibitem [{\citenamefont {Monier}\ \emph {et~al.}(2020)\citenamefont {Monier},
  \citenamefont {Huerre}, \citenamefont {Josserand},\ and\ \citenamefont
  {S{\'e}on}}]{Monier2020}%
  \BibitemOpen
  \bibfield  {author} {\bibinfo {author} {\bibfnamefont {A.}~\bibnamefont
  {Monier}}, \bibinfo {author} {\bibfnamefont {A.}~\bibnamefont {Huerre}},
  \bibinfo {author} {\bibfnamefont {C.}~\bibnamefont {Josserand}},\ and\
  \bibinfo {author} {\bibfnamefont {T.}~\bibnamefont {S{\'e}on}},\ }\bibfield
  {title} {\bibinfo {title} {Freezing a rivulet},\ }\href
  {https://doi.org/10.1103/PhysRevFluids.5.062301} {\bibfield  {journal}
  {\bibinfo  {journal} {Physical Review Fluids}\ }\textbf {\bibinfo {volume}
  {5}},\ \bibinfo {pages} {062301} (\bibinfo {year} {2020})}\BibitemShut
  {NoStop}%
\bibitem [{\citenamefont {Seguy}\ \emph {et~al.}(2023)\citenamefont {Seguy},
  \citenamefont {Protiere},\ and\ \citenamefont {Huerre}}]{Seguy2023a}%
  \BibitemOpen
  \bibfield  {author} {\bibinfo {author} {\bibfnamefont {L.}~\bibnamefont
  {Seguy}}, \bibinfo {author} {\bibfnamefont {S.}~\bibnamefont {Protiere}},\
  and\ \bibinfo {author} {\bibfnamefont {A.}~\bibnamefont {Huerre}},\
  }\bibfield  {title} {\bibinfo {title} {Role of geometry and adhesion in
  droplet freezing dynamics},\ }\href
  {https://doi.org/10.1103/PhysRevFluids.8.033601} {\bibfield  {journal}
  {\bibinfo  {journal} {Physical Review Fluids}\ }\textbf {\bibinfo {volume}
  {8}},\ \bibinfo {pages} {033601} (\bibinfo {year} {2023})}\BibitemShut
  {NoStop}%
\bibitem [{\citenamefont {Bumma}\ \emph {et~al.}(2023)\citenamefont {Bumma},
  \citenamefont {Huerre}, \citenamefont {Pierre},\ and\ \citenamefont
  {S{\'e}on}}]{Bumma2023}%
  \BibitemOpen
  \bibfield  {author} {\bibinfo {author} {\bibfnamefont {K.}~\bibnamefont
  {Bumma}}, \bibinfo {author} {\bibfnamefont {A.}~\bibnamefont {Huerre}},
  \bibinfo {author} {\bibfnamefont {J.}~\bibnamefont {Pierre}},\ and\ \bibinfo
  {author} {\bibfnamefont {T.}~\bibnamefont {S{\'e}on}},\ }\bibfield  {title}
  {\bibinfo {title} {Early freezing dynamics of an aqueous foam},\ }\href
  {https://doi.org/10.1039/D3SM00278K} {\bibfield  {journal} {\bibinfo
  {journal} {Soft Matter}\ ,\ \bibinfo {pages} {10.1039.D3SM00278K}} (\bibinfo
  {year} {2023})}\BibitemShut {NoStop}%
\bibitem [{\citenamefont {Mullins}\ and\ \citenamefont
  {Sekerka}()}]{Mullins1964}%
  \BibitemOpen
  \bibfield  {author} {\bibinfo {author} {\bibfnamefont {W.~W.}\ \bibnamefont
  {Mullins}}\ and\ \bibinfo {author} {\bibfnamefont {R.~F.}\ \bibnamefont
  {Sekerka}},\ }\bibfield  {title} {\bibinfo {title} {Stability of a {{Planar
  Interface During Solidification}} of a {{Dilute Binary Alloy}}},\ }\href
  {https://doi.org/10.1063/1.1713333} {\ \textbf {\bibinfo {volume} {35}},\
  \bibinfo {pages} {444}}\BibitemShut {NoStop}%
\bibitem [{\citenamefont {Langer}()}]{Langer1980}%
  \BibitemOpen
  \bibfield  {author} {\bibinfo {author} {\bibfnamefont {J.~S.}\ \bibnamefont
  {Langer}},\ }\bibfield  {title} {\bibinfo {title} {Instabilities and pattern
  formation in crystal growth},\ }\href
  {https://doi.org/10.1103/RevModPhys.52.1} {\ \textbf {\bibinfo {volume}
  {52}},\ \bibinfo {pages} {1}}\BibitemShut {NoStop}%
\bibitem [{\citenamefont {Worster}()}]{Worster1986}%
  \BibitemOpen
  \bibfield  {author} {\bibinfo {author} {\bibfnamefont {M.~G.}\ \bibnamefont
  {Worster}},\ }\bibfield  {title} {\bibinfo {title} {Solidification of an
  alloy from a cooled boundary},\ }\href
  {https://doi.org/10.1017/S0022112086002938} {\ \textbf {\bibinfo {volume}
  {167}},\ \bibinfo {pages} {481}}\BibitemShut {NoStop}%
\bibitem [{\citenamefont {Huppert}()}]{Huppert1990}%
  \BibitemOpen
  \bibfield  {author} {\bibinfo {author} {\bibfnamefont {H.~E.}\ \bibnamefont
  {Huppert}},\ }\bibfield  {title} {\bibinfo {title} {The fluid mechanics of
  solidification},\ }\href {https://doi.org/10.1017/S0022112090001938} {\
  \textbf {\bibinfo {volume} {212}},\ \bibinfo {pages} {209}}\BibitemShut
  {NoStop}%
\bibitem [{\citenamefont {Banin}\ and\ \citenamefont {Anderson}()}]{Banin1974}%
  \BibitemOpen
  \bibfield  {author} {\bibinfo {author} {\bibfnamefont {A.}~\bibnamefont
  {Banin}}\ and\ \bibinfo {author} {\bibfnamefont {D.~M.}\ \bibnamefont
  {Anderson}},\ }\bibfield  {title} {\bibinfo {title} {Effects of {{Salt
  Concentration Changes During Freezing}} on the {{Unfrozen Water Content}} of
  {{Porous Materials}}},\ }\href {https://doi.org/10.1029/WR010i001p00124} {\
  \textbf {\bibinfo {volume} {10}},\ \bibinfo {pages} {124}}\BibitemShut
  {NoStop}%
\bibitem [{\citenamefont {Fikiin}()}]{Fikiin1998}%
  \BibitemOpen
  \bibfield  {author} {\bibinfo {author} {\bibfnamefont {K.~A.}\ \bibnamefont
  {Fikiin}},\ }\bibfield  {title} {\bibinfo {title} {Ice content prediction
  methods during food freezing: A survey of the {{Eastern European}}
  literature},\ }\href {https://doi.org/10.1016/S0260-8774(98)00120-4} {\
  \textbf {\bibinfo {volume} {38}},\ \bibinfo {pages} {331}}\BibitemShut
  {NoStop}%
\bibitem [{\citenamefont {Succar}\ and\ \citenamefont
  {Hayakawa}()}]{Succar1990}%
  \BibitemOpen
  \bibfield  {author} {\bibinfo {author} {\bibfnamefont {J.}~\bibnamefont
  {Succar}}\ and\ \bibinfo {author} {\bibfnamefont {K.}~\bibnamefont
  {Hayakawa}},\ }\bibfield  {title} {\bibinfo {title} {A {{Method}} to
  {{Determine Initial Freezing Point}} of {{Foods}}},\ }\href
  {https://doi.org/10.1111/j.1365-2621.1990.tb03606.x} {\ \textbf {\bibinfo
  {volume} {55}},\ \bibinfo {pages} {1711}}\BibitemShut {NoStop}%
\bibitem [{\citenamefont {Rahman}()}]{Rahman2009}%
  \BibitemOpen
  \bibfield  {author} {\bibinfo {author} {\bibfnamefont {M.~S.}\ \bibnamefont
  {Rahman}},\ }\href@noop {} {\emph {\bibinfo {title} {Food {{Properties
  Handbook}}}}}\ (\bibinfo  {publisher} {CRC Press})\ \Eprint
  {https://arxiv.org/abs/F9FoNy06qvcC} {F9FoNy06qvcC} \BibitemShut {NoStop}%
\bibitem [{\citenamefont {Caccavo}\ and\ \citenamefont
  {Lamberti}()}]{Caccavo2017}%
  \BibitemOpen
  \bibfield  {author} {\bibinfo {author} {\bibfnamefont {D.}~\bibnamefont
  {Caccavo}}\ and\ \bibinfo {author} {\bibfnamefont {G.}~\bibnamefont
  {Lamberti}},\ }\bibfield  {title} {\bibinfo {title} {{{PoroViscoElastic}}
  model to describe hydrogels' behavior},\ }\href
  {https://doi.org/10.1016/j.msec.2017.02.155} {\ \textbf {\bibinfo {volume}
  {76}},\ \bibinfo {pages} {102}}\BibitemShut {NoStop}%
\bibitem [{\citenamefont {Caccavo}\ \emph {et~al.}()\citenamefont {Caccavo},
  \citenamefont {Cascone}, \citenamefont {Lamberti},\ and\ \citenamefont
  {Barba}}]{Caccavo2018}%
  \BibitemOpen
  \bibfield  {author} {\bibinfo {author} {\bibfnamefont {D.}~\bibnamefont
  {Caccavo}}, \bibinfo {author} {\bibfnamefont {S.}~\bibnamefont {Cascone}},
  \bibinfo {author} {\bibfnamefont {G.}~\bibnamefont {Lamberti}},\ and\
  \bibinfo {author} {\bibfnamefont {A.~A.}\ \bibnamefont {Barba}},\ }\bibfield
  {title} {\bibinfo {title} {Hydrogels: Experimental characterization and
  mathematical modelling of their mechanical and diffusive behaviour},\ }\href
  {https://doi.org/10.1039/C7CS00638A} {\ \textbf {\bibinfo {volume} {47}},\
  \bibinfo {pages} {2357}}\BibitemShut {NoStop}%
\bibitem [{\citenamefont {Hu}\ and\ \citenamefont {Suo}()}]{Hu2012}%
  \BibitemOpen
  \bibfield  {author} {\bibinfo {author} {\bibfnamefont {Y.}~\bibnamefont
  {Hu}}\ and\ \bibinfo {author} {\bibfnamefont {Z.}~\bibnamefont {Suo}},\
  }\bibfield  {title} {\bibinfo {title} {Viscoelasticity and poroelasticity in
  elastomeric gels},\ }\href {https://doi.org/10.1016/S0894-9166(12)60039-1} {\
  \textbf {\bibinfo {volume} {25}},\ \bibinfo {pages} {441}}\BibitemShut
  {NoStop}%
\bibitem [{\citenamefont {Davies}\ \emph {et~al.}()\citenamefont {Davies},
  \citenamefont {Huang}, \citenamefont {Harper}, \citenamefont {Hook},
  \citenamefont {Thomas}, \citenamefont {Burgar},\ and\ \citenamefont
  {Lillford}}]{Davies2010}%
  \BibitemOpen
  \bibfield  {author} {\bibinfo {author} {\bibfnamefont {E.}~\bibnamefont
  {Davies}}, \bibinfo {author} {\bibfnamefont {Y.}~\bibnamefont {Huang}},
  \bibinfo {author} {\bibfnamefont {J.~B.}\ \bibnamefont {Harper}}, \bibinfo
  {author} {\bibfnamefont {J.~M.}\ \bibnamefont {Hook}}, \bibinfo {author}
  {\bibfnamefont {D.~S.}\ \bibnamefont {Thomas}}, \bibinfo {author}
  {\bibfnamefont {I.~M.}\ \bibnamefont {Burgar}},\ and\ \bibinfo {author}
  {\bibfnamefont {P.~J.}\ \bibnamefont {Lillford}},\ }\bibfield  {title}
  {\bibinfo {title} {Dynamics of water in agar gels studied using low and high
  resolution {{1H NMR}} spectroscopy},\ }\href
  {https://doi.org/10.1111/j.1365-2621.2010.02448.x} {\ \textbf {\bibinfo
  {volume} {45}},\ \bibinfo {pages} {2502}}\BibitemShut {NoStop}%
\bibitem [{\citenamefont {Rafsanjani}\ \emph {et~al.}()\citenamefont
  {Rafsanjani}, \citenamefont {Jin}, \citenamefont {Deng},\ and\ \citenamefont
  {Bertoldi}}]{Rafsanjani2019}%
  \BibitemOpen
  \bibfield  {author} {\bibinfo {author} {\bibfnamefont {A.}~\bibnamefont
  {Rafsanjani}}, \bibinfo {author} {\bibfnamefont {L.}~\bibnamefont {Jin}},
  \bibinfo {author} {\bibfnamefont {B.}~\bibnamefont {Deng}},\ and\ \bibinfo
  {author} {\bibfnamefont {K.}~\bibnamefont {Bertoldi}},\ }\bibfield  {title}
  {\bibinfo {title} {Propagation of pop ups in kirigami shells},\ }\href
  {https://doi.org/10.1073/pnas.1817763116} {\ \textbf {\bibinfo {volume}
  {116}},\ \bibinfo {pages} {8200}}\BibitemShut {NoStop}%
\bibitem [{\citenamefont {Stiti}\ \emph {et~al.}()\citenamefont {Stiti},
  \citenamefont {Castanet}, \citenamefont {Labergue},\ and\ \citenamefont
  {Lemoine}}]{Stiti2020a}%
  \BibitemOpen
  \bibfield  {author} {\bibinfo {author} {\bibfnamefont {M.}~\bibnamefont
  {Stiti}}, \bibinfo {author} {\bibfnamefont {G.}~\bibnamefont {Castanet}},
  \bibinfo {author} {\bibfnamefont {A.}~\bibnamefont {Labergue}},\ and\
  \bibinfo {author} {\bibfnamefont {F.}~\bibnamefont {Lemoine}},\ }\bibfield
  {title} {\bibinfo {title} {Icing of a droplet deposited onto a subcooled
  surface},\ }\href {https://doi.org/10.1016/j.ijheatmasstransfer.2020.120116}
  {\ \textbf {\bibinfo {volume} {159}},\ \bibinfo {pages} {120116}}\BibitemShut
  {NoStop}%
\end{thebibliography}%

\end{document}